# Building Short Value Chains for Animal Welfare-Friendly Products Adoption: Insights from a Restaurant-Based Study in Japan

Takuya Washio[1], Sota Takagi[1], Miki Saijo[1], Ken Wako[3], Keitaro Sato[1], Hiroyuki Ito[1], Ken-ichi Takeda[2], and Takumi Ohashi[1*]

[1]Institute of Science Tokyo, 2-12-1 Ookayama Meguro-ku, Tokyo 152-8550, Japan
* ohashi.t.f540@m.isct.ac.jp
[2]Institute of Agriculture, Academic Assembly, Shinshu University, Kamiina, Nagano 399-459, Japan
[3]EBI Marketing Co., Ltd., Tokyo, Japan

**Abstract**
As global attention on sustainable and ethical food systems grows, animal welfare-friendly products (AWFP) are increasingly recognized as essential to addressing consumer and producer concerns. However, traditional research often neglects the interdependencies between production, retail, and consumption stages within the supply chain. This study examined how cross-stage interactions among producers, consumers, and retail intermediaries can promote AWFP adoption. By establishing a short value chain from production to consumption, we conducted a two-month choice experiment in the operational restaurant, employing a mixed-method approach to quantitatively and qualitatively assess stakeholder responses. The results revealed that providing information about AWFP practices significantly influenced consumer behavior, increasing both product selection and perceived value. Retailers recognized the potential for economic benefits and strengthened customer loyalty, while producers identified new revenue opportunities by re-fattening delivered cow. These coordinated changes—defined as synchronized actions and mutual reinforcement across production, retail, and consumption—generated positive feedback loops that motivated stakeholders to adopt AWFP practices. This research underscores the potential of strategically designed short value chain to foster cross-stage coordination and highlights their role as practical entry points for promoting sustainable and ethical food systems on a larger scale.

1 **Introduction**

For the stable supply of food and the sustainability of livestock farming, which is the primary method globally, the importance of animal-welfare-friendly livestock farming is increasing worldwide. In response, research is actively being conducted today across the entire value chain, including the development and dissemination of production practices based on animal behavior science (Gregory 2008; Grandin 2010; Schwartzkopf-Genswein et al. 2012; Aguayo-Ulloa et al. 2014; Del Campo et al. 2014), the establishment of certification and quality assurance systems and regulations (Wood, Holder, and Main 1998; Jacques 2014), and consumer education and acceptance promotion (Grunert, Bredahl, and Brunsø 2004; Montossi et al. 2013; Sonoda et al. 2018; Mohan, Maheswarappa, and Banerjee 2022; Washio et al. 2023).

However, in Japan, it has been noted that the adoption of animal welfare is lagging compared to Western countries, where the concept was adopted early (Shiga et al. 2020; Washio, Ohashi, and Saijo 2020; Toyota and Tan 2024). Among consumers in Japan, the recognition rate of animal welfare is low (Takeda et al. 2010; Shiga et al. 2020; Washio, Ohashi, and Saijo 2020), and there is no widely adopted national certification. Factors contributing to this delay include the fact that the concept was imported from the West, and Japan is currently responding to international pressure (Amos 2022; Eurogroup for Animals 2022). Therefore, in the Japanese market, which acts as a follower, it is important not only to expect changes based on the market principle where businesses respond to consumer demand but also to promote transformation from both production and consumption perspectives.

Research on the societal implementation of animal-welfare-friendly products (AWFP) faces methodological constraints due to the lack of comprehensive studies observing the entire value chain from production to consumption. Traditional research has primarily analyzed individual stages of the supply chain based on the premise of mass production and consumption of food and livestock products. This fragmented approach limits the insights to specific stages of the value chain, posing challenges for achieving a holistic transformation that incorporates AWFP principles.

Studies on animal welfare within the supply chain have focused on independent stages, assuming scenarios where producers operate mainly through auctions or wholesale markets, with minimal direct interaction with consumers, investigating factors that promote animal welfare practices (Fraser 2008; Veissier et al. 2008). Additionally, numerous studies have examined consumer preferences and acceptance, particularly in the context of purchasing activities in supermarkets (Sonoda et al. 2018; Apostolidis and McLeay 2019; Díaz-Caro et al. 2019; Czine et al. 2020; Cornish et al. 2020; Washio et al. 2023). While these studies provide valuable insights optimized for specific stages of the value chain, they also highlight the need for changes in other stages, raising a "chicken and egg" problem regarding which stage should change first.

For instance, previous research indicates that producers are more likely to adopt animal welfare practices when there are economic benefits (Nocella, Hubbard, and Scarpa 2010). However, for this to happen, there needs to be a demand for AWFP, which is challenging for producers to initiate. Similarly, consumers are more inclined to purchase AWFP if the prices are affordable, which requires increased production and price reductions at the retail stage, both of which are different stages of the value chain.

Therefore, to broadly implement AWFP in society, it is necessary for the entire value chain to transform simultaneously, with interactions among actors across different stages promoting or hindering practices at each stage. Considering the mutual influence between producers and consumers is crucial to breaking the current status quo. Thus, it is essential to explore opportunities arising from positive interactions experienced simultaneously by actors at different stages of the value chain.

This study aims to elucidate how cross-stage interactions, which traditional research has overlooked, promote animal welfare practices among producers, consumers, and retail intermediaries. By forming a minimal value chain from production to consumption in real-world settings where AWFP production, retail, and consumption occur, this research explored opportunities that lead to the promotion of animal welfare practices experienced by producers, retail intermediaries, and consumers. We conducted a choice experiment in the operating restaurant for two months and observed offering of AWFP influence them in mixed method approach both quantitatively and qualitatively. The objective of this study is to derive opportunities to advance animal welfare practices by comprehensively observe the entire AWFP value chain, thereby promoting simultaneous transformation across the value chain and providing practical insights for the societal implementation of AWFP.

## 2 Material and methods

In this study, we conducted action research centered on an experiment where dishes made with animal-welfare-friendly beef were served in the operating restaurant. Through this experiment, we employed a mixed-method research approach to observe the changes experienced by producers, distributors, retailers, and consumers involved in the animal-welfare-friendly product value chain from both quantitative and qualitative perspectives. Action research (Hellier et al. 2003; Wittmayer et al. 2014), a participatory and democratic approach that simultaneously instigates change and improvement in practices. Action research is collaborative in nature, often being undertaken by teams comprising both researchers and practitioners (Erro-Garcés and Alfaro-Tanco 2020). Its strength lies in its active inclusion of practitioners, not merely as subjects, but as engaged participants in the research process. When applied to the context of farm animal welfare, this collaboration could result in a deeper, multifaceted understanding of opportunities to advance animal welfare practices.

Animal-welfare-friendly products, for end consumers, are products where the accompanying information adds value. Therefore, comparing the presence or absence of information is considered effective in contrasting animal-welfare-friendly products with traditional livestock products. Thus, a key part of the experiment involved comparing consumer reactions with and without the provision of information.

2.1  Settings and context

The experiment was conducted in the restaurant located in a campus of a university in Greater Tokyo area. The restaurant is owned by a private company with business in several industrial fields. The restaurant is operated by the employees of the restaurant division, one was the chef led the operation and 2 to 5 servers according to the hourly busyness. Before the experiment, the manager needed practices to enhance its business sustainability and joined the current experiment including a test sales of animal-welfare friendly beef meals. The Japanese Black beef with its brand name "Kuroshima-Kuroushi" was provided by a producer in Kuroshima, Okinawa Prefecture, Japan, who incorporated with universities in development of animal-welfare friendly beef cattle production system. The producer fattened a Japanese Black cow (aged 13 years) that was due to be culled after the breeding period, under intensive feeding for 3 months, and then sold the carcass. Until it was fattened, this cow was pastured all day except when fed. The restaurant had about 60 seats and opened from 11 a.m. to 4 p.m., serving three kinds of lunch menu of the day from 11:30 a.m. to 3:30 p.m., and beverages and desserts for entire opening hour. The restaurant was in a self-service style where customers made orders at the cashier and receive the tickets. Once the meal is ready, customers were called to the service counter and exchange their tickets with the meal.

2.2  Information presentation design

The information presentation was designed in cooperation with a professional designer with experience in advertising consumer goods and other products. The following five elements were incorporated into the information presentation based on the attributes that the producer wanted to promote, the desired expressions for display during business hours at the restaurant, the designer's suggestions, and expected effects on customers' behavior based on previous studies.

**Farm animal welfare (AW)**

As awareness of the term of animal welfare and its meanings in Japan was expected to be low (Washio, Ohashi, and Saijo 2019; Takeda et al. 2010), it was critical to clearly explain them to customers. Previous researches suggested that providing detailed information about animal welfare such as on-package labels increased intention to purchase over conventional welfare products (Lagerkvist and Hess 2011; Alonso, González-Montaña, and Lomillos 2020). In this study, in addition to explaining the term animal welfare, which is mental and physical state of an animal from birth to death in according with the definition of WOAH (WOAH, n.d.), the text information expressed that the beef provided was raised with consideration for animal welfare.

**Producer (PD)**

Producer information behind the farming products such as practices caring animal welfare and environmental sustainability (Miyama and Morita 2023) or challenges that producers facing (Scozzafava et al. 2020) were suggested to encourage consumers to choose those products by increasing willingness to pay. In this study, the textual information expressed that the producer is committed to keeping cows in a more natural way and is working on pasturage, and that it is conducting demonstration experiments with a university to promote farm animal welfare in Japan, and incorporated images of the producers.

**Place of origin (PO)**

Agricultural product information of place of origin was suggested to influence consumers' perception and preferences for the products (Scozzafava et al. 2020). Especially when it is local origin, it was suggested to increase consumers' perceived product quality (Kumpulainen et al. 2018). Japanese consumers were suggested to prefer Japanese produced meat (Washio et al. 2023). In this study, the fact that the beef is domestically produced, that it was produced on Kuroshima, a remote island in Okinawa Prefecture, and that Kuroshima is a suitable environment for grazing are expressed as textual information, and an image depicting the location of Kuroshima was incorporated.

**Pastured (PS)**

Being year-round pastured was one of the key practice concerning animal welfare of Kuroshima-Kuroushi beef offered in this study. Previous study suggested the presence of consumer segment who value pastured beef (Schulze, Spiller, and Risius 2021). In this study, taking advantage of the topography of the production area, the textual information expressed that the cattle are raised on pasture, which the producer is particular about, and that Kuroshima is a suitable environment for grazing, and incorporates images of grazing cattle.

**Creditability (CR)**

Information behind the products needs credibility. Previous studies suggest that creditability characteristics influence consumers' expectations and willingness to pay on animal-welfare-friendly products (Napolitano, Girolami, and Braghieri 2010; Cornish et al. 2020). In this study, the textual information expressed the fact that the farm was assessed by an ethologist (Co-author) as having guaranteed animal welfare in accordance with the Shinshu Comfort Livestock Production Accreditation Standard (Nagano Prefecture 2007).

The media of choice were posters next to the menu and education cards. The posters (Figure 1) were displayed next to the menu inside/outside the restaurant during the weeks when the information presentation was present. In addition, customers who ordered a meal using Kuroshima-Kuroushi were offered an education card (Figure 2) on a tray with the meal.

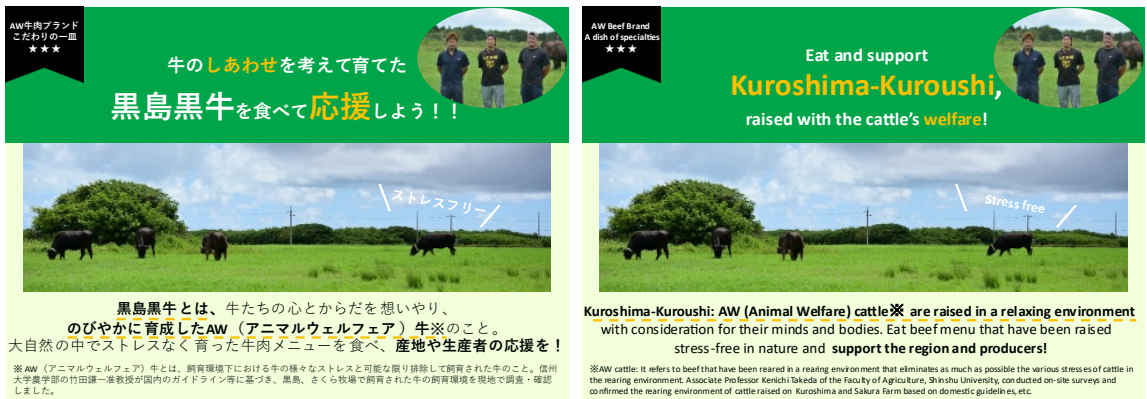

Figure 1 The poster presented during the week with information presentation (The actual size of the card was 148 mm X 210 mm, left: original version in Japanese; right: English translation)

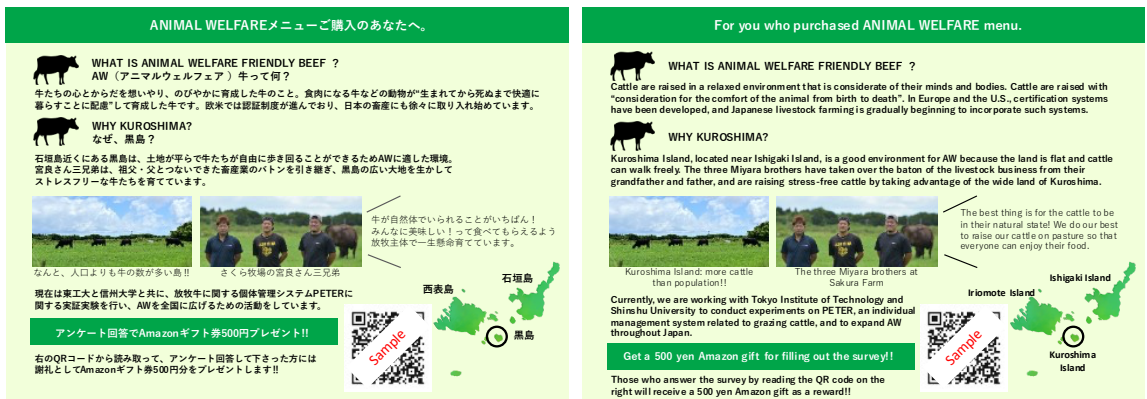

Figure 2 The cards offered during the week with information presentation (The actual size of the card was 74 mm X 105 mm, left: original version in Japanese; right: English translation)

2.3 Data collection
2.3.1 Kuroshima-Kuroushi meal sales in the operating restaurant
2.3.1.1 Preparation of Kuroshima-Kuroushi

The beef served in the experiment was prepared by the producer. His present business model was primarily a ranching operation focused on producing and selling Japanese Black calves on year-round pasture. In a typical lifecycle, one cow would produce approximately seven calves prior to retirement. Upon retirement, these cows were traditionally sold to butchers at auctions. However, the producer has opted for an alternative approach by attempting to re-fatten delivered cows. The goal was to market this as a higher-quality, animal-welfare-friendly beef, given that these cows have spent most of their lives grazing on a single farm, which inherently promotes better animal welfare. Accordingly, one such cow was re-fattened, slaughtered, and butchered by the producer, then the beef was delivered directly to the restaurant. This approach not only potentially increased the product's value, but it also imbued the process with a unique narrative that respects the animal's life and well-being, a factor that could be of interest to discerning consumers.

2.3.1.2 Pricing between the producer and the restaurant

The procurement price of the beef was determined through a negotiation between the producer and the restaurant ensuring a feasible condition that would extend beyond the confines of experimental control. Therefore, in this study, restaurants play the role of both distributors and retailers. The producer's breakeven point was established at 1500 JPY per kilogram, a figure arrived at after considering the costs of re-fattening, slaughtering, butchering, and shipping. Conversely, the restaurant had a breakeven point of 2000 JPY per kilogram to ensure their ability to serve a meal at their standard pricing structure. After detailed discussions, both parties arrived at a mutually agreeable price of 1800 JPY per kilogram. Thus, within this specific context, the economically efficient procurement of animal-welfare-friendly beef was not only realized but also empirically demonstrated. This negotiation process highlights the potential for the successful integration of ethical considerations into the economic framework of food production and consumption.

2.3.1.3 Offering at the restaurant

Three types of menus (roasted-meat bowl, stew, and steak) used the beef was developed by the chef and served at the restaurant for 6 weeks on the fixed two days in a week. Figure 3 and 4 illustrate how information was presented during the offering period. The menus served on each day is presented in Table 1. The information presentation with the posters and the cards was present and absent bi-weekly. In order to assess the effects of information presentation on customers' menu choice, evaluation of the restaurant, and value recognition on Kuroshima-Kuroushi, several data was collected throughout the period.

Table 1 Menus served during the experiment

| Date | Day | Menu 1 Ingredient | Dish | Menu 2 Ingredient | Dish | Menu 3 Ingredient | Dish |
|---|---|---|---|---|---|---|---|
| 13-Sep | Tue | BEEF | Roasted-meat bowl | Pork | Stew | Chicken | Curry |
| 15-Sep | Thu | Pork | Deep fires | BEEF | Stew | Chicken | Curry |
| 20-Sep | Tue | BEEF | Roasted-meat bowl | Chicken | Stew | Chicken | Curry |
| 22-Sep | Thu | Beef | Steak | BEEF | Stew | Chicken | Curry |
| 27-Sep | Tue | BEEF | Roasted-meat bowl | Pork | Stew | Chicken | Curry |
| 29-Sep | Thu | BEEF | Stew | Chicken | Deep fries | Chicken | Curry |
| 4-Oct | Tue | BEEF | Roasted-meat bowl | Lamb | Stew | Chicken | Curry |
| 6-Oct | Thu | Pork | Deep fires | BEEF | Stew | Beef | Hashed beef |
| 11-Oct | Tue | Chicken | Fried-chicken bowl | BEEF | Stew | Chicken | Curry |
| 14-Oct | Fri | BEEF | Steak | Pork | Stew | Chicken | Curry |
| 18-Oct | Tue | Pork | Roasted-meat bowl | BEEF | Stew | Chicken | Curry |
| 21-Oct | Fri | BEEF | Steak | Pork | Stew | Chicken | Curry |

BEEF refers to Kuroshima-Kuroushi.
Shading means that there was information presentation.

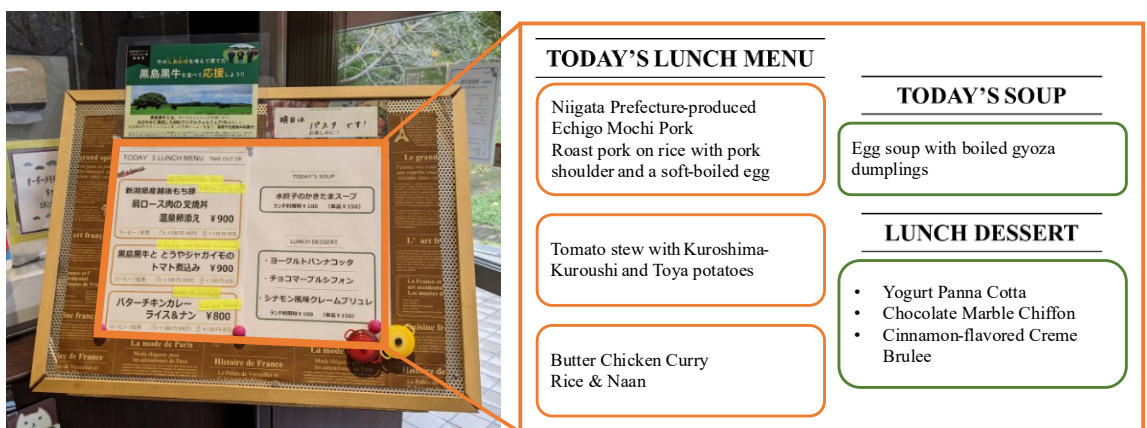

Figure 3 The poster presented at the cashier (Photo taken during the experiment)

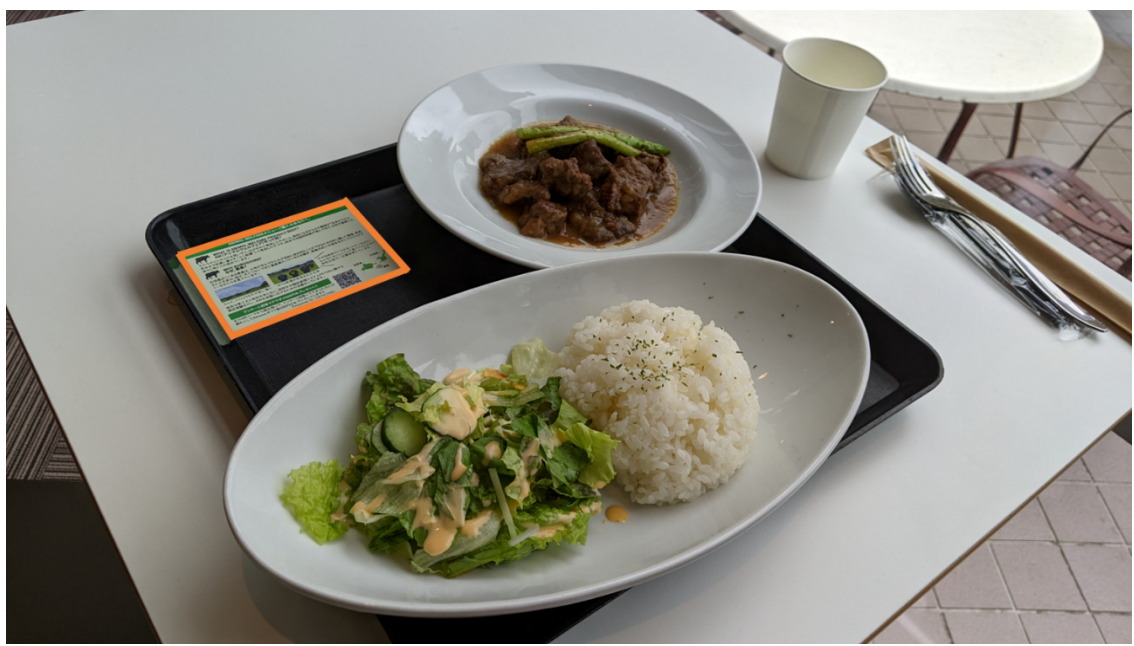

Figure 4 The card offered with Kuroshima-Kuroushi stew on the tray (Photo taken during the experiment)

The effects of information presentation were evaluated based on three questions. 1. If the information presentation affected on customers' lunch meal choice; 2. If the information presentation affected on customers' evaluation of the restaurant; 3. If the information presentation affected on customers' value recognition of Kuroshima-Kurosushi. To evaluate the effects of information presentation, both quantitative and qualitative data were collected and analyzed. Quantitative data were obtained through the number of sales and an online questionnaire accessible from the QR code printed on the education card.

2.3.2 Customer questionnaire

The questionnaire was designed with three parts: respondent screening, evaluation on the restaurant, and value recognition of Kuroshima-Kuroushi (Table 2). To enable quantitative comparison between the presence and absence of the information presentation, a card with QR code containing the same link to the online questionnaire (Figure 2 and 5) were offered on the tray with Kuroshima-Kuroushi meal during the week with and without information presentation. Customers who responded to the questionnaire received a reward in the form of a money certificate worthed 500 JPY. Each education card had a printed unique response ID to restrict the number of responses to one for one card. In order to limit multiple responses from the same person chasing rewards, the rewards were given in exchange for the declaration of an e-mail address, of which only one is issued uniquely to each student, faculty, and staff member by the university located on the campus of the restaurant. Throughout the period, the questionnaire response rate was 18.4%, and 126 responses were collected.

Table 2 Three parts consisted of the online questionnaire

| Part | Purpose |
| --- | --- |
| Respondent screening | To assess if the respondent is eligible to answer |
| Restaurant evaluation | To evaluate if the information presentation affects customers' evaluation of the restaurants |
| Value recognition of Kuroshima-Kuroushi | To evaluate if the information presentation affects customers' value recognition of Kuroshima-Kuroushi |

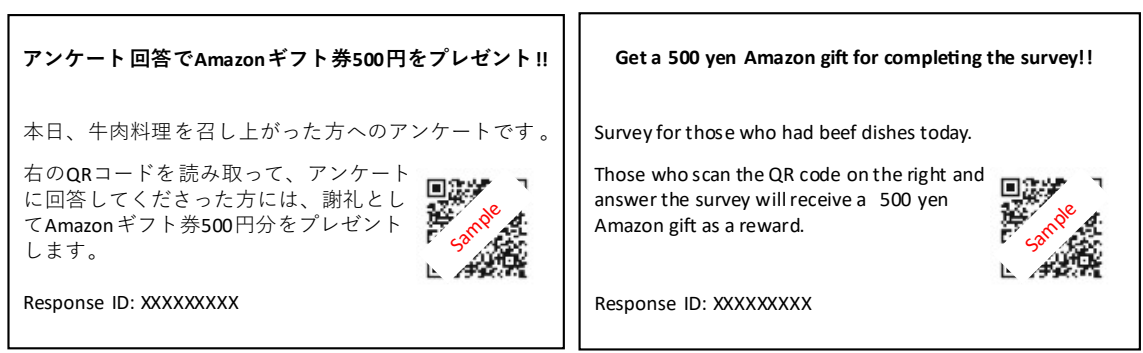

Figure 5 Card offered during the week without information presentation (The actual size of the card was 55 mm X 91 mm, left: original version in Japanese; right: English translation)

2.3.2.1 Customers' evaluation of the restaurant

A customer's evaluation of the restaurant is composed of a variety of factors. In this experiment, we focused on the following five factors.

**Satisfaction (SF)**

The fulfillment of customer expectations for the restaurant is one of the most primary evaluations a customer gives to the restaurant. Customer satisfaction is described as "a judgment that a product or service feature, or the product or service itself, provided (or is providing) a pleasurable level of consumption-related fulfillment, including levels of under or over fulfillment" (Oliver 2010).

**Perceived food quality (FQ)**

Customers' perception of the quality of food served in the restaurant influences their evaluation of the restaurant. In addition to intrinsic quality, which depends on the physical condition of the food, such as taste, aroma, appearance, and shape, extrinsic quality, such as price, brand, origin, and production information, can also be evaluated (Oude Ophuis and Van Trijp 1995)

**Perceived value (PV)**

The value of the experience, such as food and service, for the price paid at the restaurant influences the customer's evaluation of the restaurant. Perceived value has been noted to have a significant impact on consumer purchasing behavior (Wang 2015). Perceived value is described as "the customer's overall appraisal of the net worth of the service, based on the customer's assessment of what is received (benefits provided by the service), and what is given (costs or sacrifice in acquiring and utilizing the service)" (Hellier et al. 2003).

**Revisit intention (RI)**

One of the most important evaluations a customer gives to the restaurant is whether or not he or she will want to return after finishing a meal at the restaurant. Revisit intention is described as "an affirmed likelihood to revisit the restaurant in both the absence and presence of a positive attitude toward the provider" (Han, Back, and Barrett 2009).

**Word of mouth (WM)**
Based on their experience at the restaurant, customers' evaluations of it to other people, including family and friends, are evaluations that reflect the actions they subsequently take toward the restaurant. Word of mouth is described as "person-to-person, oral communication between a communicator and receiver which is perceived as a non-commercial message" (Arndt 1967).

Previously validated scales were employed for this study. Konuk measured the influence of perceived food quality, price fairness, perceived value and satisfaction on the organic food restaurant customers' revisit intention and word-of-mouth intentions (Konuk 2019). Scales for satisfaction, food quality, perceived value, revisit intention and word of mouth were employed for this study. Items were modified to align the context of the study and measured with seven-point Likert scales (1 = strongly disagree, 7 = strongly agree). The questionnaire items used is shown in Table 3.

Table 3 Questionnaire items to measure customers' evaluation of the restaurant

| Category | Item |
| --- | --- |
| Satisfaction (SF) | |
| | I am satisfied with my decision to visit this restaurant. |
| | My choice of this restaurant was a wise decision. |
| | I am glad I made the decision to visit this restaurant. |
| Food Quality (FQ) | |
| | The meal served looked delicious. |
| | The meal provided was healthy. |
| | The meal provided was delicious. |
| | The meal provided was made with fresh ingredients |
| Perceived Value (PV) | |
| | This restaurant served meal that was worth the price. |
| | Overall, the meal at this restaurant was worth it. |
| | The experience at this restaurant was worth the price. |
| Revisit Intention (RI) | |
| | I would visit this restaurant again. |
| | I would eat a meal at this restaurant again. |
| | I would consider using this restaurant again. |
| Word of Mouth (WM) | |
| | I would recommend this restaurant if asked for advice. |
| | I would refer this restaurant positively to anyone I know. |
| | I would encourage other people to visit this restaurant. |

2.3.2.2 Effects on customers' value recognition

A scale was developed for each of the five elements included in the information presentation to measure respondents' perception of value. Three items were created for each information elements: impressions of the information presented, perceived value of the credence attributes of beef, and willingness to support activities related to the production of beef with credence attributes. The items are shown in Table 4. Each item was measured with seven-point Likert scales (1 = strongly disagree, 7 = strongly agree). Table 7 shows the questionnaire items used.

Table 4 Questionnaire items to measure customers' value recognition on Kuroshima-Kuroushi

| Category | Item |
| --- | --- |
| Animal-welfare Friendly Farming | |
| | I think the impressions of animal-welfare friendly farming are good. |
| | I want to support and assist animal-welfare friendly farming. |
| | I consider animal-welfare friendly farming to be necessary or valuable to me. |
| Producer | |
| | I think the impressions of the producers of the beef served are good. |
| | I want to support and assist the producer of the beef served |
| | I consider the producer of the beef served to be necessary or valuable to me. |
| Place of Origin | |
| | I think the impressions of the place of origin of the beef served are good. |
| | I want to support and assist the place of origin of the beef served. |
| | I consider the place of origin of the beef served to be necessary or valuable to me. |
| Pastured beef | |
| | I think the impressions of the pasture-raised beef are good. |
| | I want to support and assist the pasture-raised beef. |
| | I consider the pasture-raised beef to be necessary or valuable to me. |
| Appraisal | |
| | I think the impressions of the appraisal of the beef served are good. |
| | I want to support and assist in the appraisal of the beef served. |
| | I consider the appraisal of the beef served to be necessary or valuable to me. |

2.3.3 Stakeholder interviews

2.3.3.1 Customer depth-interview

The interviewees were selected from customers who ordered the Kuroshima-Kuroushi menu items during the weeks with information presentation. The co-authors explained the purpose of the study to these customers after their meal and requested their participation in the interview. After obtaining consent, interviews were conducted with 13 individuals. Semi-structured interviews were expected to get subjective reflections of the experience at the restaurant which we may have not been anticipated. An interview guide was developed putting the focus on exploring customers' evaluation on their experiences in the restaurant and the impressions and reflections of the served beef and information presentation. The following contents were included in the interview guide. 1. Participant background information; 2. Impressions and reflection of the served beef meal; 3. Impression and reflection of the information presentation. Each interview lasted around 15 minutes. Each session was audio recorded with the participants' consent and transcribed for analysis. Customers who responded to the questionnaire or participated in the interview received a reward in the form of a money certificate worthed 500 JPY.

2.3.3.2 Focus group discussion with the restaurant chef and staff

To reflect the process of the field restaurant experiment from the viewpoint of the restaurant, a focus group interview using a semi-structured interview was conducted targeting the chef and the staff at the restaurant, after the field restaurant experiment was completed. The results from the customers' meal choice, restaurant evaluation, and value recognition were presented and briefly explained taking around 15 mins. After the explanation, a semi-structured interview was conducted. Three main questions were prepared as follows; 1.What were the good things that happened to you during the experiment? 2. What hard things happened to you during the experiment? 3. How do you personally think about animal-welfare friendly beef?

2.3.3.3 Producer depth interview

To reflect the process of the field restaurant experiment from the viewpoint of the producer of the beef, a semi-structured interview was conducted online, after the field restaurant experiment was completed. The results from the customers' meal choice, restaurant evaluation, value recognition, and the restaurant's reflection of the sales activity were presented and briefly explained taking around 15 mins. After the explanation, a semi-structured interview was conducted. Three main questions were prepared as follows; 1. How was the impression to the experiment? 2.What were hard things happened to you during the experiment? 3.How do you think about future re-fattening of cows for animal-welfare-friendly beef production?

2.4 Data analysis

Our data analysis consists of five distinct phases: 1. information presentation effects on customers' lunch meal choice, 2. customers' evaluation of the restaurant, and 3. customers' value recognition of Kuroshima-Kurosushi. 4. evaluation of the offering case from the restaurant's point of view 5. evaluation of the offering case from the producer point of view. Statistical analysis was conducted with R 4.0.3 (R Core Team 2020).

2.4.1 Information presentation effect on customers' lunch meal choice

To assess the effect of information presentation on the customers' Kuroshima-Kuroushi meal selection, logistic regression was applied. It analyzed the effects of information presentation, type of meal that Kuroshima-Kuroushi was cooked, available alternative lunch menu other than those using Kuroshima-Kuroushi. For information presentation, one dummy variable was prepared (1 donated present). For the type of meal that Kuroshima-Kuroushi was cooked, three dummy variables of steak, roasted beef bowl, and stew was prepared. For those three Interaction terms were prepared for these three independent variables and the information presentation dummy. For available alternatives, three dummy variables of pork stew, chicken curry, and pork fries – alternatives appeared more than twice during the experiment – were prepared. The model took those as independent variables, and consumer's selection of Kuroshima-Kuroushi meal (also a dummy variable) as dependent variable. Additionally, we will provide qualitative data from open-ended responses in questionnaires and interviews with customers to help interpret the quantitative data for the readers.

2.4.2 Information presentation effect on customers' evaluation of the restaurant

A confirmatory factor analysis was conducted using lavvan package (Rosseel 2012) to convert questionnaire responses to scores. To evaluate whether the measurement model fits the data well, fit indexes of RMSAE (MacCallum, Browne, and Sugawara 1996) and CFI (Hu and Bentler 1999) were assessed. Using the calculated factor scores, Welch's t-test was conducted to assess the differences between the periods when information presentation was present/absent. Among the factors whose score difference between the periods were statistically significant, Cohen's d (Cohen 1988; 1992) was calculated to assess the effect size. Additionally, we will provide qualitative data from open-ended responses in questionnaires and interviews with customers to help interpret the quantitative data for the readers.

2.4.3 Information presentation effect on customers' value recognition of Kuroshima-Kuroushi

An exploratory factor analysis was conducted to extract factors that each item loads. Items which were found to be loaded to the intended factors were processed to a confirmatory factor analysis using lavvan package (Rosseel 2012) to validate the measurement model and to convert questionnaire responses to scores. Using the calculated factor scores, Welch's t-test was conducted to assess the differences between the periods when information presentation were present/absent. Among the factors whose score difference between the periods were statistically significant, Cohen's d (Cohen 1988; 1992) was calculated to assess the effect size. Additionally, we will provide qualitative data from open-ended responses in questionnaires and interviews with customers to help interpret the quantitative data for the readers.

2.4.4 Evaluation of the offering case from the restaurant's point of view
Based on the transcripts collected from the focus group discussion, we describe the restaurant's evaluation of the offering case in a narrative approach. The narrative approach is regarded as an effective method to grasp the complex dynamics of people's thoughts and behaviors by using vivid examples from research observations (Sandelowski 1994).

2.4.5 Evaluation of the offering case from the producer point of view
Based on the transcripts collected from the in-depth interviews, we narratively present the producer's evaluation of the offering case. The narrative approach is considered an effective method for capturing the complex dynamics of individuals' thoughts and behaviors, utilizing vivid examples from research observations (Sandelowski 1994).

# 3 Result
## 3.1 Customers' meal choice Information presentation effects on customers' lunch meal choice

Selection rate of the Kuroshima-Kuroushi meal was calculated for each date and presented in Figure 6. The rate was calculated using the number of lunch menu served on each day while all the three alternatives were available.

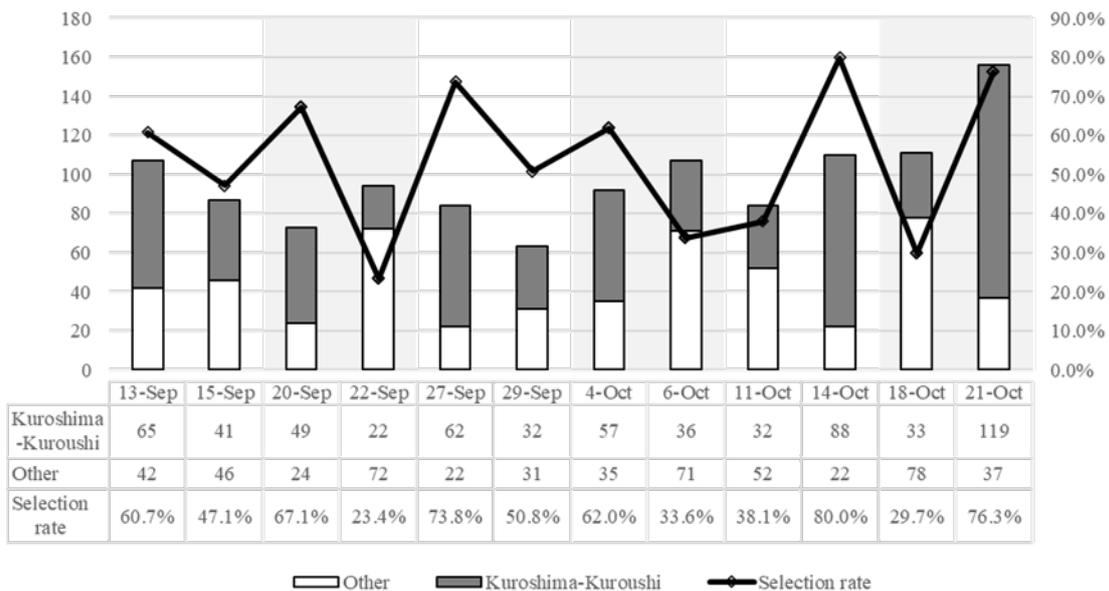

Figure 6 Transition of Kuroshima-Kuroushi menu selection rate. Numbers below each date in the table refer to the count of ordered lunch meals using Kuroshima-Kuroushi and the sum of the other two alternatives. Selection rate refers to the share of Kuroshima-Kuroushi menu in all the lunch meal order counts on each day. Shaded dates denote the presence of the information presentation. Kuroshima-Kuroushi meal served on each date is shown in Table 1.

From the logistic regression, it was found that, holding all other predictor variables constant, the odds of customer's choice of Kuroshima-Kuroushi menu occurring increased by 69% (95% CI [1.10, 2.59]) for the presence of information presentation, increased by 139% (95% CI [1.24, 4.61]) for the presence of Kuroshima-Kuroushi steak in the alternative , increased by 91% (95% CI [1.03, 3.54]) for the presence of pork stew in the alternative, and, increased by 251% (95% CI [1.68, 7.32]) for the presence of chicken curry in the alternative. Among independent variables, stew dummy and interaction term between stew and information presentation was excluded from the model due to multicollinearity. The results are presented in Table 5.

Table 5 Results of binary logit regression analysis

|  | B | S.E. | Wald | p | Odds Ratio | 95% C.I.for OR Lower | Upper |
|---|---|---|---|---|---|---|---|
| (Intercept) | -1.515 | 0.473 | 10.263 | 0.001 | 0.220 | | |
| Information presentation | 0.525 | 0.218 | 5.797 | 0.016 | 1.690 | 1.103 | 2.591 |
| Steak | 0.871 | 0.336 | 6.723 | 0.010 | 2.390 | 1.237 | 4.617 |
| Roasted beef bowl | 0.315 | 0.218 | 2.078 | 0.149 | 1.370 | 0.893 | 2.102 |
| Pork stew as alternative | 0.646 | 0.315 | 4.212 | 0.040 | 1.908 | 1.029 | 3.537 |
| Chicken curry as alternative | 1.255 | 0.375 | 11.200 | 0.001 | 3.509 | 1.682 | 7.320 |
| Pork fries as alternative | 0.311 | 0.281 | 1.226 | 0.268 | 1.365 | 0.787 | 2.368 |
| Presentation: Steak | -0.614 | 0.353 | 3.031 | 0.082 | 0.541 | 0.271 | 1.080 |
| Presentation: Roasted beef bowl | 0.315 | 0.218 | 2.078 | 0.149 | 1.370 | 0.893 | 2.102 |

Nagelkerke R2: 0.103, Hosmer and Lemeshow goodness of fit (GOF) test: Chi-squared = 0.47624, df = 8, p-value = 0.803
Shaded rows were statistically significant.

The customer interviews revealed insightful results on the factors influencing the choice of animal-welfare-friendly meals. Three key elements were identified that swayed customers towards these meals: the type of dish, the place of origin of the beef, and the novelty factor associated with the meal.

*The type of dish or ingredients played a significant role in guiding customers' meal selections.*

"I ordered it because I wanted roast beef, but I was a little curious about the product, like from Okinawa."

"I was interested in this kind of display at the cash register, and I also simply like meat."

Additionally, the provenance of the beef, in this case, originating from Okinawa, also appeared to influence the decision-making process.

*"I noticed something that is not always there. Animal-welfare-friendly beef, it's Okinawan beef, isn't it?"*

The novelty aspect, gleaned from the credence cues provided, was perceived as a compelling factor, making the animal-welfare-friendly meal more appealing to the customers.

*"It's fun to have a menu that you don't usually get to eat."*

*"I looked at the menu list and thought, where was it produced, it looks different from usual. Was it from Okinawa? I think it says so."*

Interestingly, it was also suggested that customers who were knowledgeable about animal welfare issues found the animal-welfare-friendly beef meal particularly enticing.

*"I noticed these (posters) on the menu today. I answered a questionnaire about animal welfare the other day and thought, "Oh, they are doing that kind of work." The image is good."*

*"If you see a sign like that when you go to the supermarket, you will be a little more aware of it. I think that if you go to a supermarket and see a sign like that, you might become a little more aware of it. In supermarkets, where there are many different kinds of products on sale, I am sure that the words "animal welfare" will be lost in the stream of information."*

These results clearly demonstrate that providing information about animal welfare significantly influences consumer decision-making.

3.2  Effects on customers' evaluation of the restaurant

The result of confirmatory factor analysis is presented in Table 6. For food quality, average variance extracted was below the recommended threshold of 0.5, however we considered the convergent validity of the construct is still adequate as composite reliability was higher than 0.6 (Fornell and Larcker 1981).

Table 6 Scale items for the restaurant evaluation, convergent validity and reliability

| Category | Item | Estimate | α | CR | AVE |
|---|---|---|---|---|---|
| Satisfaction (SF) | | | 0.87 | 0.88 | 0.71 |
| | I am satisfied with my decision to visit this restaurant. . | 0.85 | | | |
| | My choice to choose this restaurant is a wise one. | 0.86 | | | |
| | I am happy about my decision to visit this restaurant | 0.82 | | | |
| Food Quality (FQ) | | | 0.76 | 0.73 | 0.45 |
| | Food presentation was visually attractive. | 0.60 | | | |
| | The restaurant offered healthy food | 0.57 | | | |
| | The restaurant served tasty food. | 0.69 | | | |
| | The restaurant provided fresh food. | 0.79 | | | |
| Perceived Value (PV) | | | 0.82 | 0.84 | 0.62 |
| | Food served in this restaurant was a good value for the price. | 0.80 | | | |
| | The overall value of eating food was high. | 0.74 | | | |
| | The food was worth the money. | 0.83 | | | |
| Revisit Intention (RI) | | | 0.91 | 0.91 | 0.77 |
| | I will keep visiting this food restaurant in the future. | 0.92 | | | |
| | I would like to come back to the restaurant in the future. | 0.91 | | | |
| | I will consider revisiting the restaurant in the future | 0.80 | | | |
| Word of Mouth (WM) | | | 0.87 | 0.87 | 0.68 |
| | I will recommend this restaurant to other people who seek my advice. | 0.81 | | | |
| | I will say positive things to my acquaintances about this restaurant. | 0.86 | | | |
| | I will encourage other people to visit this restaurant. | 0.83 | | | |

Measurement Model Fit Indexes: $\chi2 = 118.412$, df = 80, p value = 0.003, GFI = 0.883, AGFI = 0.824, NFI = 0.920, CFI = 0.972 (well fit), RMSAE = 0.062 (mediocre fit), AIC = 3579.393
CR refers to composite reliability; AVE denotes average variance extracted.

Using the calculated scores, the differences between scores measured in the period with information presentation and without were assessed. Welch t-test showed that the differences were statistically significant for satisfaction, food quality. With Cohen's d (Cohen 1988; 1992), the effect size in satisfaction was small, and in food quality was medium. The result is presented in Table 7.

Table 7 Welch t-test for customers' evaluation of the restaurant

| | Without information (N = 35) | | With information (N = 90) | | | | | 95% CI | | |
|---|---|---|---|---|---|---|---|---|---|---|
| | Mean | S.D. | Mean | S.D. | t | P | df | Lower | Higher | Cohen's d |
| SF | -0.082 | 0.739 | 0.210 | 0.767 | -1.972 | 0.049 | 59.870 | -0.621 | -0.016 | 0.396 |
| FQ | -0.069 | 0.464 | 0.177 | 0.492 | -2.551 | 0.013 | 57.988 | -0.438 | -0.053 | 0.517 |
| PV | -0.073 | 0.693 | 0.189 | 0.708 | -1.877 | 0.065 | 61.109 | -0.542 | 0.017 | |
| RV | -0.063 | 0.652 | 0.163 | 0.673 | -1.698 | 0.095 | 60.195 | -0.493 | 0.040 | |
| WM | -0.071 | 0.724 | 0.182 | 0.735 | -1.740 | 0.086 | 61.159 | -0.545 | 0.038 | |

Welch Two sample t-test, two-sided.
Mean1 denotes mean scores measured during the period with absence of information presentation whereas Mean 2 denotes that with presence of it.

From the results of the t-test, a significant increase in SF and FQ with the information presentation was suggested. On the other hand, there were no significant differences in PV, RV, and WM. This may be due to the characteristics of the restaurant where experiment was conducted. First, the frequency of visits by the respondents indicates that most of them visit the restaurant repeatedly. Since there are only two restaurants operating on campus, and there are few restaurants in the surrounding area, there is not an abundance of restaurant options for customers, and this situation is likely to result in a large number of repeat customers.

During the interviews, some of the customers said,

*"The food was good as usual."*

And

*"I often come to this restaurant because it offers good value for money for what I eat."*

Given these facts, it can be assumed that the restaurant has always given its regular customers a high level of satisfaction with its prices and has maintained a good reputation and willingness to revisit, and that these were not affected by the information presentation.

The results gleaned from the customer interviews shed light on various perceptions towards the offering of animal-welfare-friendly beef at the restaurant. A segment of customers indicated that they found the quality of the animal-welfare-friendly beef served to be superior. This is evident from comments like,

*"Animal-welfare-friendly beef roasted beef bowl, the meat was very tasty and went well with the sautéed mushrooms."*

Furthermore, several customers expressed their intention to revisit the restaurant specifically for the animal-welfare-friendly beef menu. For instance, one customer stated,

*"I used to come to the restaurant for lamb or steak days, but the black beef I had this time was so good that I would like to come back when black beef is on the menu,"*

demonstrates the positive impact of the animal-welfare-friendly beef menu on customer loyalty. Importantly, customers also voiced a sense of value and satisfaction from their restaurant visit. Statements such as,

*"It is nice to have food of this quality at this price on campus,"*

*"Good value for money and good food,"*

And

*"The food was more than the price,"*

emphasize that the customers found the animal-welfare-friendly beef menu to offer good value for money.

In summary, the animal-welfare-friendly beef menu was viewed favorably by customers, who generally expressed positive feedback about their experience at the restaurant. This underlines the potential for eateries to enhance their reputation and customer loyalty by incorporating animal-welfare-friendly options into their menus.

### 3.3 Effects on customers' value recognition

Exploratory factor analysis was conducted to extract factors that each item loads. The result of each item's loadings on the constructs is presented in Table 8.

Table 8 Scale items measuring customers' value recognition on Kuroshima-Kuroushi and reliability

| Item | Animal-welfare friendly farming | Pastured beef | α |
|---|---|---|---|
| I think the impressions of animal-welfare friendly farming are good. | 0.89 | | 0.87 |
| I consider animal-welfare friendly farming to be necessary or valuable to me. | 0.63 | | |
| I want to support and assist animal-welfare friendly farming. | 0.51 | | |
| I think the impressions of the pasture-raised beef are good. | | 0.77 | 0.83 |
| I consider the pasture-raised beef to be necessary or valuable to me. | | 0.76 | |
| I want to support and assist the production of pasture-raised beef. | | 0.75 | |

Kaiser–Meyer–Olkin measure of Sampling Adequacy: Overall MSA = 0.85.
Bartlett's Test of Sphericity: ChiSq = 2,824.068, p = 0.000.
Extraction method: Minimum residual with oblique rotation.
RMSEA = 0.0722.
Factor loadings < |0.3| are not shown.

Two factors were extracted and named according to the intended measurements. As only items targeting the value recognition of animal-welfare friendly farming and pastured beef were found to load on the intended factors, these two were processed to a confirmatory factor analysis. The result of the confirmatory factor analysis is presented in Table 9. Factor scores were calculated for the following analysis.

Table 9 Scale items measuring customers' value recognition on Kuroshima-Kuroushi, convergent validity and reliability

| Category | Item | Loadings | α | CR | AVE |
|---|---|---|---|---|---|
| Animal-welfare Friendly Farming | | | 0.87 | 0.88 | 0.70 |
| | I think the impressions of animal-welfare friendly farming are good. | 0.87 | | | |
| | I want to support and assist animal-welfare friendly farming. | 0.78 | | | |
| | I consider animal-welfare friendly farming to be necessary or valuable to me. | 0.85 | | | |
| Pastured beef | | | 0.83 | 0.83 | 0.62 |
| | I think the impressions of the pasture-raised beef are good. | 0.81 | | | |
| | I want to support and assist the pasture-raised beef. | 0.77 | | | |
| | I consider the pasture-raised beef to be necessary or valuable to me. | 0.78 | | | |

Measurement Model Fit Indexes: $\chi^2$ = 36.449, df = 8, p value = 0.000, GFI = 0.93, AGFI = 0.81, NFI = 0.91, CFI = 0.93 (well fit), RMSAE = 0.069 (well fit), AIC = 1939.289
CR denotes composite reliability, AVE denotes Average Variance Extracted.

Using the calculated scores, the differences between scores measured in the period with information presentation and without were assessed. Welch t-test showed that the differences were statistically significant for animal welfare, producer, place of origin, pastured, creditability. With Cohen's d (Cohen 1988; 1992), the effect size in producer, place of origin, pastured, creditability were medium, and in animal welfare was large. The result is presented in Table 10.

Table 10 Welch t-test for customers' value recognition on Kuroshima-Kuroushi

| | Without information (N = 35) | | With information (N = 90) | | | | | 95% CI | | |
|---|---|---|---|---|---|---|---|---|---|---|
| | Mean 1 | S.D. | Mean 2 | S.D. | t | P | df | Lower | Higher | Cohen's d |
| AW | -0.196 | 0.867 | 0.504 | 0.931 | -3.849 | 0.000 | 58.282 | -1.064 | -0.336 | 0.779 |
| PS | -0.141 | 0.674 | 0.363 | 0.922 | -2.942 | 0.004 | 48.796 | -0.848 | -0.160 | 0.624 |

Welch Two sample t-test, two-sided
Mean 1 denotes mean scores measured during the period with absence of information presentation whereas Mean 2 denotes that with presence of it.

A significant increase in the two elements with the information presentation was suggested. In particular, in relation to AW, which had the largest effect size of the differences, the interviewees' statements revealed many responses that they first learned about farm animal welfare through exposure to the information provided during the experiment.

In addition, the customer interviews yielded results highlighting various sentiments associated with the consumption of animal-welfare-friendly beef. Some customers expressed a sense of safety as a perceived value derived from consuming animal-welfare-friendly beef. One participant shared,

*"I guess this was done with the cows in mind, but I felt that it would be safe for the eaters as well. It was such good meat. It was very tasty."*

This underscores the connection made between animal welfare practices and the quality and safety of the food.

Another set of customers felt empathy towards the producers of the beef. The reality of the challenges associated with rearing cows without causing stress, often highlighted in media, resonated with them. Comments such as,

> "There are a lot of them these days, aren't there? Cows and raising them without stress. I have seen it on TV. It must be hard work. They must be grazing their cows, right?"

And

> "I don't have a negative image of it, but I thought that typhoons and such would be difficult,"

reflect this empathy.

Moreover, there was a group of customers who expressed a willingness to support the initiative of animal-welfare-friendly production. Responses like

> "I think your approach to AW is very good and I look forward to seeing quality ingredients adopted into your menu."

And

> "It is difficult to do, but I wish I could support such efforts."

as well as

> "It seems like efficiency is not going to increase much and it will be hard work, but I would like to support it."

elucidate this supportive sentiment.

Thus, it can be concluded that effective information presentation infuses the act of "eating beef" with additional layers of contextual value, making consumers consider aspects such as the welfare of the cows, the producers' efforts, and the larger initiative towards animal welfare. This further reinforces the power of informed consumption in driving sustainable and ethical food choices.

3.4 Evaluation of the offering case from the restaurant's point of view

First, we asked the chef and the staff to freely reflect on what they felt was good and what they felt was difficult about their efforts during the experiment from their perspectives. First, the staff in charge of customer service reflected on the positive reception he received from customers through the exchange of questions and responses from customers.

> "The voices from the customers. Many of them call out to us and ask us things like, 'Is this a different meat than usual?' or 'Is this a special meat?' People who just drink coffee, or sometimes eat lunch, they call out to us while looking at the menu list." (the staff)

> "The people who asked me the question were quite positive." (the staff)

It was suggested that some customers perceived the offering of Kuroshima-Kuroushi along with the information as positive. Furthermore, it suggests that staff also reflected back on the customers' positive reactions as favorable reactions for the restaurant business.

Next, the chef in charge of menu development and preparation, who described himself as "not a very negative thinker," offered positive comments about the beef he used. We was concerned about the negative impact on meat quality, price, and logistics since this was the first time the producer had shipped cattle that had been re-fattened and since the beef was delivered directly from the producers. However, contrary to our concerns, we heard some positive reflections. First, the quality of the meat was comparable to other beef that he had purchased in the past.

> "We know that the meat quality is also good, so I thought it would be about the same as usual heifer. To be frank, there is nothing special about them because they are stress-free. I think it's just a nuance of what you would expect from Wagyu heifer beef." (the chef)

Regarding price, it was noted that even after taking into account the cost of preparation, such as the removal of unnecessary parts, and other labor costs, the beef was still less expensive than beef purchased pre-processed.

> "So the cost of the meat is 1800 yen per kilo this time, but with the heifer beef we were using, a sirloin costs 3000 yen per kilo or so. But, well, that is also a bit expensive because the fat is trimmed off as part of the yield. Wagyu beef is 7,000 or 8,000 if it was a sirloin as well." (the chef)

After clarifying that, due to the nature of its location within the university, the restaurant has secured profits by keeping its selling price constant and reducing its cost of food sold, he suggested that the current delivery price is favorable to the restaurant. In addition, he was eager to consider developing a high value-added menu if there was an improvement in customers' evaluation of the restaurant, which Kuroshima-Kuroushi offerings was suggested to have improved.

> "Hmmm. For example, there are customers who are particular about their food, like vegetarians. When I was working at a teppan, there were a lot of them. They had a strong individuality, or they

*said, "I am this way." If there are customers like that, I have to go to them. Socially, too, such a tendency has increased over the past few years. (omission) I think these people are more likely to show interest. They will eat the meat even if it is a little expensive. They will eat the meat even if it is not at a low cost because they are told that it is made with a particular method." (the chef)*

*"For events, for example. For example, we charge a little more for Christmas. We put foie gras on it. But it's not every day. But it's not every day. We have to try it out once, sometimes special; otherwise, people might just show up without thinking and leave if they think it's too expensive." (the chef)*

Thus, the chef's experience suggests that Kuroshima-Kuroushi, appeared to be something that could attract the interest of people who prefer higher value-added meals with a story behind the production process. There was also a reflection that the nature of the restaurant was suitable, which made it easier to incorporate beef with niche characteristics.

*"In this restaurant, that is not so much the case. I knew how much to serve, whether it was steak or stew, because all the customers, students or otherwise, came to the restaurant. I knew how much I was going to serve, whether it was steak or stew, for example. I could read how much I was going to serve every day. But if you think about it in a regular restaurant, it would go up and down depending on the number of people coming in and the reaction of the customers, so with this kind of food, it would be very difficult. It is very difficult in such places. Here, it's somewhat constant, so I can predict the amount of food that will be sold this week. But in a general restaurant, you can't do that, so in that sense, I think it is difficult." (the chef)*

The fact that the restaurant in the experiment had a large number of customers and a limited number of items suggests that the sales volume was predictable, allowing the producer to deliver a certain amount of beef and the restaurant to sell the entire amount of beef, an activity that was possible. This suggests a troubling situation for producers who are trying to produce animal-welfare friendly beef. In contrast to this restaurant, restaurants that take the form of upscale establishments that offer services tailored to the particulars and tastes of a small number of customers who can be expected to sell the product as a high value-added product suggest that it is more difficult to be a stable seller because the volatility in sales volume is not always high. In encouraging beef producers to engage in animal-welfare friendly production, it is desirable to resolve this contradiction and accumulate knowledge from the perspective of achieving stable sales volume and high added value in restaurants.

Finally, we asked about their personal evaluation of the animal-welfare friendly beef production. It was noted that there have been changes throughout this experiment.

*"Me? I think it's the human ego from the very beginning. I thought, "Either way, you raise them for meat and kill them. When I heard this from [A], I thought, "That's just a twist in words. Conversely, we are in the position of cooking and selling the meat, so receiving their lives is necessary no matter what. Well, that's the image I have of it, and it's not so much a good thing or a bad thing." (the chef)*

*(Interviewer: When there is animal welfare meat and non animal welfare meat, you mentioned the unit price earlier, if the response from customers was high, or if there were other factors, would it be favorable even if it was ego?)*

*"Yes, I agree. I think it is necessary to respond to what customers want." (the chef)*

*"When I heard the term "animal welfare," I thought it was the same thing, that the cows would end up being eaten or being eaten. But after talking with customers and thinking about it, I realized that it is better for cows to live happily while they are alive because they are eventually eaten, but we should appreciate their lives." (the staff)*

It was suggested that both the chef and the staff experienced changes in personal impressions of animal-welfare friendly products. Both the chef and the employee recalled having critical impressions of animal-welfare friendly products prior to their participation in the field restaurant experiment. They described them as being more expensive, leading to increased costs for the restaurant, and as an unnecessary activity of the human ego. Their impressions of the value of animal-welfare friendly products were suggested to have been enhanced through the experiment period via dialogue in response to questions from customers and their own interpretations of product introductions possibly contribute the restaurant's business. This suggests that employees' proactive participation in activities to understand customer perceptions and products, such as improving information provision measures, may play an important role not only in improving the measures, but also in improving employee perceptions.

3.5 Evaluation of the offering case from the producer point of view

First, we asked him to freely reflect on what they felt was good and what they felt was difficult about their efforts for the experiment and the outcomes from their perspectives. First, the producer reflected the general impression positive and fun and showed the relieved emotion.

*"I'm glad that we did it. Honestly I was worried that the quality of the meat would be worse. I wondered if it would taste good. I just fed them because I don't have any fattening skills; I fed them for 5-6 months."*

*"I'm glad to hear that the beef tastes good even though it's a delivered cow. But I am sure we can make it even tastier." (the producer)*

Thus, the producer offered a positive reflection on the experiment. There was a sense of relief and satisfaction, not only because the quality of the meat from delivered cows met his expectations but also due to the positive response from customers. He found the process enjoyable and felt confident about enhancing the taste even further in the future.

Then we asked the producer to reflect what he found difficult or to be improved during the activity.

*"Cows that are no longer pregnant are either put out of service, auctioned, or re-fattened. When they are auctioned, some of the most expensive cows fetch 300,000-400,000 yen. It is important that the weight of the cattle is large, and cattle weighing more than 600 kg are considered expensive. Nowadays, the market price has dropped, so cheap cows can cost as little as 100,000 yen.*

*It is the easiest for producers to sell their cattle at auctions because the buyers take care of the slaughtering and butchering. In this experiment, the hardest part was to send the cattle to be butchered and then shipped. It costs a little over 100,000 yen to slaughter and pack them in blocks. It costs about 100,000 for six months of fattening and additional feed. Next time, I would like to sell a whole cow if possible." (the producer)*

The producer shed light on the economic dimensions of the livestock industry when the topic turned to the difficulties he faced. He compared the cost-efficiency of traditional auctioning, where cows could fetch up to 400,000 yen depending on their weight, against the expenses incurred in our experiment. The costs for butchering, packaging and, shipping, and feeding were found to be significant, prompting him to desire to sell a whole cow for future opportunities.

Finally, we asked the producer to tell what he was thinking to do in the future, reflecting the experiment.

*"It is motivating to know that there is a demand and value for meat from re-fattened aged beef. It is also more enjoyable than selling at the auction at a low price of 100,000."*

*"I would definitely do it again. It was fun. I am sure that from now on, we will be able to sell our products even if they are expensive, and the term "animal welfare" is often seen in the newspapers. Consumers will be able to eat more comfortably. I am sure that we will be able to sell our products at a higher price than at auctions. If we were to do it, we would like to be able to sell them at a higher price than at the auction. It is not profitable to do it little by little, so eventually, it would be easier and less time-consuming to do it in batches of five or ten animals. Eventually, I would like to be able to sell them as Kuroshima local brand beef. It will take a long time, but in order to do that, I think it will be necessary to feed the cows produced on Kuroshima Island." (the producer)*

Though this study, the producer expressed an optimistic future outlook, drawing inspiration from the demand for meat from re-fattened aged beef. Despite the higher costs, the prospect of better prices and a growing consciousness around animal welfare among consumers made the endeavor worthwhile. He envisages scaling up his operations, potentially selling in larger batches, and establishing Kuroshima local brand beef. To attain this goal, he identified a critical step in the path: utilizing feeds produced locally on Kuroshima Island.

4 Discussion

In this study, we designed a value chain for AWFP through a case study of sales activities in real-world settings, providing insights from four perspectives to promote AWFP adoption. These opportunities suggest that simultaneous changes in production, retail, and consumption can lead to mutual interactions among players in various stages of the supply chain, fostering a willingness to adopt AWFP practices.

### 4.1 Consumer Awareness and Behavior

Previous research has indicated that consumers recognize both intrinsic values such as taste (Thorslund et al. 2016; Alonso, González-Montaña, and Lomillos 2020; Humble, Palmér, and Hansson 2021) and extrinsic values such as the story behind the product (Napolitano, Girolami, and Braghieri 2010; van Riemsdijk et al. 2023), which increases their willingness to purchase and support AWFP. Our study confirmed that consumers recognize extrinsic values related to the product's story, not just intrinsic values, and become more active in purchasing and consuming AWFP. This suggests that actively communicating the production practices, vision, and efforts of producers can enhance consumer behavior by appealing to their self-identity and the context of support through consumption activities. Our findings demonstrate that consumer recognition of the producer's background, previously suggested through hypothetical experiments, can indeed promote consumer behavior in more complex real-world environments. This can potentially create a cycle where consumers become more active in purchasing AWFP, resulting in economic benefits for producers, distributors and retailers, thereby promoting further AWFP adoption. Understanding and supporting the producer's vision and efforts can expand consumer choices and foster sustainable consumption patterns.

### 4.2 Producer Motivation and Behavior

Producers' recognition of positive responses from consumers and the restaurant, as well as the resulting sales volume and economic benefits, suggests an increased willingness to adopt animal-welfare-friendly practices. Various claims have been made regarding whether animal welfare-oriented production practices can be profitable for producers. On the cost side, there is a possibility of increased costs due to the need for more resources in rearing (Fernandes et al. 2021; Bessei 2018), while others argue that improved productivity can reduce costs and enhance efficiency (Støier et al. 2016; Dawkins 2017). On the revenue side, higher meat quality can add value (Napolitano, Girolami, and Braghieri 2010; Miranda-de la Lama et al. 2017), and new added value from the background of AW-oriented production can improve profitability (Nocella, Hubbard, and Scarpa 2010; Janssen, Rödiger, and Hamm 2016; Sonoda et al. 2018; Washio et al. 2023). Our study confirmed that consumers recognized not only the intrinsic value but also the extrinsic value related to the product's story, leading to increased consumption and purchasing. Additionally, the case study suggested the feasibility of creating new revenue opportunities for producers by utilizing unused resources. The beef provided in this study was from cows that had been delivered and were re-fattened, turning a previously nearly worthless asset into a high-value product. This motivated producers to engage in AWFP production, allowing them to enjoy economic benefits and potentially enhancing the sustainability of AW-oriented production methods. Recognizing new revenue opportunities for producers can promote AWFP adoption and expand animal welfare-oriented production methods.

### 4.3 Retail Motivation and Behavior

Retails recognize the positive consumer responses, sales volume, and economic rationality, increasing their willingness to handle AWFP and develop proactive strategies for business success. The motivation for retailers to handle AWFP has been attributed to economic/business benefits (Schulze, Spiller, and Risius 2019; Jones and Comfort 2022) and ethical values (Schulze, Spiller, and Risius 2019). In Japan, where the AWFP market is still developing and economic benefits are unclear, existing advanced initiatives are likely driven by ethical values. In this case, the chef and the staff at the restaurant who initially had a negative perception of AWFP changed their view to a positive one through suggestions of business opportunities and interactions with consumers. Specifically, they recognized the economic rationality of handling AWFP through positive consumer reactions and became more willing to actively handle AW-oriented products. This can help bridge the gap between producers and consumers, promoting AWFP adoption. Recognizing the economic rationality of handling AWFP and developing proactive strategies can make the expansion of the AWFP market a reality, leading to broader dissemination of animal welfare-oriented products.

### 4.4 Economic Rationality for Producers and Distributors

Producers and the restaurant recognize the economic rationality in their mutual transactions of AWFP, increasing their willingness to continue and expand these transactions. Unless producers deal directly with consumers, the economic rationality for intermediaries is essential to establish a realistic AWFP value chain (Akaichi and Revoredo-Giha 2020). This means that either lower purchase costs or higher selling prices for the target products must be achievable. In this case, the producer incurred additional costs for re-fattening delivered cows. Thus, it is necessary for the producer to receive added value that covers these additional costs. In this case, the restaurant, acting as a distributor and retailer, could purchase beef of comparable quality with added value from the production background story at a price comparable to or lower than usual and provide the food to consumers. This price was sufficient for the producer to cover the re-fattening costs and generate profit, demonstrating the potential for greater economic rationality for both parties. Previous research has suggested that the key to ensuring the economic rationality of AWFP lies in the potential to increase top-line revenue through added value passed on to consumers (Nocella, Hubbard, and Scarpa 2010; de Jonge and van Trijp 2013). However, the results of this study indicate that it is possible to realistically promote the handling of AWFP without necessarily increasing the product price. This finding is particularly significant for retailers and restaurants that are unable to adopt premium pricing strategies, as it provides a compelling incentive to handle AWFP. Nevertheless, the promotion of AWFP handling critically depends on consumers' understanding and awareness of AWFP. In this study, many participants reported learning about AW for the first time through the information provided during the experiment, suggesting that awareness of animal-welfare-friendly products is currently not widespread. Instead, it is implied that consumers became aware of the value of these products when they were presented in combination with information in a restaurant setting. Therefore, producers and distributors (including restaurants) seeking to incorporate AWFP into their businesses without premium pricing strategies must actively disseminate educational information and effectively communicate the appeal of AWFP. By doing so, they can improve consumer awareness and expand the market for AWFP within their respective areas.

### 4.5 Simultaneous changes in production, retail, and consumption

Compared to the complex value chain with many participants assumed in traditional research focused on mass production and consumption, the short value chain realized in this study has the advantage of being achievable with the agreement of a minimal number of participants. In other words, it suggests that experimentally forming short food supply chains (SFSCs) with a minimum of intermediaries and high social proximity between producers and consumers (Galli and Brunori 2013; Takagi et al. 2024; Fujisaki et al. 2025), and simultaneously bringing about changes in production, retail, and consumption, could be a promising starting point for implementing AWFP. In addition, in production and retail, this study demonstrated that economic rationality concerning the price and trading volume of beef could be achieved as a localized optimal solution. This was possible not under the universal conditions of extensive large-scale distribution, but solely between the producer who began experimental production and restaurant, which could adjust the quantity and cooking methods of meals offered to customers, and suggesting the effectiveness of SFSCs in the transition from conventional production and distribution. Furthermore, as suggested by this study, cross-stage interaction among participants with close social distance can mutually promote each participant's practices. The sufficient interaction between different practices in this process of shortening the food supply chain is thought to contribute to the creation of added value in local production systems and new market relationships(Renting, Marsden, and Banks 2003). Thus, achieving a SFSCs for AWFP on a small scale can become a positive loop. This loop provides an opportunity for each participant in the food supply chain to explore their own practices, discover added value, and create new markets.

### 4.6 Practical implications

Actors seeking economically rational transformations within the value chain should actively look for cooperative partners at other stages. For producers, finding retailers with strong customer loyalty, even for small quantities, could be an effective strategy. Direct sales to share the risk with retail players is another option. Developing optimization strategies for distribution channels and products could become a new area of focus. For retailers, securing motivated producers and involving consumers through information dissemination could be effective strategies. Investing in producers for intrinsic/extrinsic product differentiation and enhancing marketing efforts are potential options. Differentiation strategies through product optimization and collaboration with producers for product development and media/content strategy for product information dissemination are new areas for exploration. Additionally, policymakers aiming to realize the value chain should facilitate connections to support AWFP implementation. Identifying and fostering interactions among motivated players and enhancing capabilities for aligning interests and adjustments are necessary.

## 5  Conclusions

This study explored the promotion of animal welfare-friendly practices (AWFP) by investigating interactions across production, retail, and consumption stages within a short value chain. Through a real-world restaurant experiment, it became evident that providing detailed information about AWFP significantly influenced consumer behavior. Consumers not only increased their selection of AWFP products but also recognized their added value, emphasizing the importance of transparent communication and consumer education in fostering ethical consumption. Producers, on the other hand, discovered new economic opportunities by repurposing resources that were previously undervalued, such as re-fattened delivered cow. Positive feedback from consumers motivated producers to adopt AWFP practices further, highlighting the potential for innovative revenue models that align with sustainability goals. Meanwhile, retailers observed increased customer satisfaction and loyalty, recognizing both the economic and reputational benefits of incorporating AWFP into their offerings. This realization encouraged them to actively integrate AWFP into their business strategies, bridging the gap between production and consumption. The coordinated changes observed in this study—defined as synchronized actions across production, retail, and consumption stages—created a positive feedback loop that mutually reinforced stakeholders' commitment to AWFP. This demonstrates the potential of SFSCs as a practical framework for fostering sustainable practices and facilitating cross-stage collaboration. Future research can build upon the findings presented here to develop scalable and effective strategies for promoting AWFP and fostering sustainable and ethical food systems.

Nevertheless, several limitations and directions for future research must be addressed. The first is sample selection. The selection of a specific restaurant linked the results to a subset of customers, chefs, and staff, potentially limiting the generalizability of the findings. Different restaurant customer bases, culinary philosophies, and operations can significantly influence attitudes and behaviors towards AWFP practices. The second is Variability in Consumer Acceptance. There was variability in how customers accepted and assimilated the presented information, indicating significant heterogeneity within the consumer group. Future research should consider approaches that account for different customer segments to understand the impact of information presentation on consumer choices more precisely. The third is Temporal Constraints: The study's duration may have limited the ability to observe and record long-term behavioral changes among stakeholders, including consumers. Future longitudinal studies should be considered to comprehensively understand long-term changes in behavior and attitudes influenced by production, sales activities, and dining experiences related to animal welfare. The forth is Influence of Information Media: Further exploration and validation are needed on how media influences consumer behavior. A retail and restaurant policies are complex and must consider both corporate goals and practical commercial constraints. Previous research has shown that merely adjusting menu item descriptions is insufficient for generating comprehensive knowledge in this field (Schjøll and Alfnes 2017). Therefore, our findings provide valuable insights to encourage real-world verification.

We hope the replication of these results will establish a platform for verifying many questions, laying the groundwork for future research and significantly contributing to the understanding of consumer behavior related to information media and a restaurant policy.

## 6 Acknowledgements

This work was partly supported by the Japan Science and Technology Agency's Center of Innovation Program (Grant No. JPMJCE1309) and JSPS KAKENHI (Grant No. JP24K17976). This study was approved by the Human Subject Research Ethics Review Committee at Tokyo Institute of Technology (No. 2022194). We extend our gratitude to the producer for providing beef samples and participating in our interview, and to the restaurant for incorporating our experiments into their daily operations and participating in focus group discussion. We would like to express our heartfelt gratitude to the late Junko Asakawa, whose invaluable efforts made the core field experiment at the restaurant possible and enriched this collaborative endeavor. Her dedication and contributions were essential to this study.

**CRediT Authorship Statement**
**Takuya Washio**: Conceptualization, Methodology, Software, Formal analysis, Investigation Data Curation, Writing - Original Draft, Visualization, **Sota Takagi**: Writing - Review & Editing, **Miki Saijo**: Conceptualization, Methodology, Writing - Review & Editing, Supervision, Funding acquisition **Ken Wako**: Resources **Keitaro Sato**: Investigation **Hiroyuki Ito**: Writing - Review & Editing, Funding acquisition **Ken-ichi Takeda**: Formal analysis, Investigation, Writing - Review & Editing, Funding acquisition **Takumi Ohashi**: Conceptualization, Methodology, Validation, Investigation, Writing - Review & Editing, Supervision, Project administration, Funding acquisition

**Declaration of generative AI usage**
During the preparation of this work the authors used OpenAI ChatGPT in order to review spell and grammar in proofreading process. After using this tool/service, the authors reviewed and edited the content as needed and take full responsibility for the content of the publication.